\documentclass[runningheads]{svmult}
\usepackage{makeidx}   
\usepackage{graphicx}  
\usepackage{subeqnar}  
\usepackage{multicol}  
\usepackage{cropmark} 
\usepackage{physprbb}  
\renewcommand{\greeksym}[1]{{\usefont{U}{psy}{m}{n}#1}}

\begin{document}
\title*{Nuclear reactions in the Sun after SNO and KamLAND}
\toctitle{Nuclear reactions in the Sun after SNO and KamLAND}
%
%
\titlerunning{Nuclear reactions in the Sun}
%
\author{Giovanni Fiorentini
\and Barbara Ricci}

\authorrunning{G. Fiorentini et al.}
%
%
\institute{Dipartimento di Fisica, Universit\`a di Ferrara 
 and INFN-Ferrara,\\
     Via Paradiso 12
     I-44100 Ferrara, Italy}

\maketitle              

\begin{abstract}
In this brief review we discuss the possibility of studying  the solar          
interior by means of neutrinos, in the light of the enormous progress           
of  neutrino physics in the last few years. The temperature near the           
solar center can be extracted from Boron neutrino experiments as: $ T=          
(1.57 \pm 0.01)\,  10^7 \, K$. The energy production rate in the Sun from
pp chain and CNO cycle, as deduced  from neutrino measurements, agrees             
with  the observed solar luminosity  to about twenty per cent. Progress         
in extracting astrophysical information from solar neutrinos requires           
improvement in the measurements of $^3He+$ \\$^4He \rightarrow ^7Be+\gamma$
and $p+^{14}N \rightarrow ^{15}O+ \gamma$.
\end{abstract}

\section{Introduction}

Some fourty years ago John Bahcall and Raymond Davis  
started an exploration of the Sun by means of 
neutrinos \cite{gf.jnb64,gf.davis}. 
Their journey had a long detour,
originating  the so called solar neutrino puzzle: all experiments 
- performed at Homestake, Kamioka, Gran Sasso and Baksan and
exploring different portions of the solar neutrino spectrum - reported
a deficit with respect to the theoretical 
predictions. Were all the experiments wrong? Or were  
the Standard Solar Model (SSM) calculations inadequate? Or something 
 happened to neutrinos during their hundred million
km trip from  Sun to Earth?

\begin{figure}[ht]
\begin{center}
\includegraphics[width=0.75\textwidth]{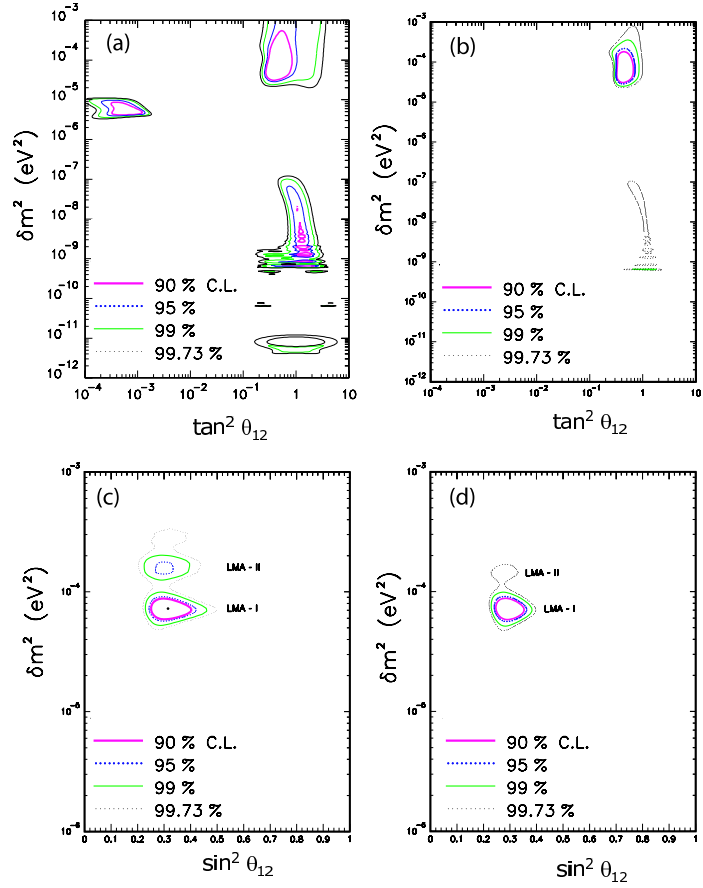}
\end{center}
\vspace{-0.5cm}
\caption[]{{\bf Global analysis of solar and terrestrial neutrino
experiments.}\\
(a) Before SNO results,
(b) including SNO-phase I, (c) including KamLAND results, (d) including
SNO-salt phase. 
From \cite{gf.fogli03,gf.foglimost,gf.fogli02}}
\label{figsno}
\vspace{-0.2cm}
\end{figure}

After thirty years the SNO experiment, 
with its  unique capability of collecting and distinguishing
events from  $\nu_e$ and from neutrinos of different flavour,   
has definitely proved that the missing electon neutrinos from the Sun  
have changed their flavour \cite{gf.sno1}.  This effect has been confirmed
by KamLAND: man made electron antineutrinos from  nuclear reactors
disappear during their few hundreds km trip to the detector
\cite{gf.kamland}. 

The enormous progress of the last few years is
summarized in Fig. \ref{figsno}. A global analysis of solar and 
reactor experiments yields for the oscillation
parameters $\delta m^2=7.1 ^{+1.2}_{-0.6} \, 10^{-5} \,eV^2$ and 
$\theta= 32.5 ^{+2.4} _{-2.3}$ degrees \cite{gf.sno2}, 
see also \cite{gf.valle,gf.smirnov,gf.fogli03}.

Really we have learnt a lot on neutrinos: their survival/transmutation
probabilities in vacuum and in matter are now substantially understood.
There is still a long road for a full description of
the neutrino mass matrix, however now that we know the fate of neutrinos
we can exploit them.

In this spirit we can go back to the original program 
started by Davis and Bahcall and ask what can be learnt on the Sun
from the study of neutrinos.
This question is clearly connected with the knowledge of 
nuclear reactions in the Sun and in the laboratory. 
Each component of the solar neutrino flux (pp, Be, B ...) 
is determined by physical and chemical properties of the
Sun (density, temperature, composition...) as well as 
by the cross sections of the pertinent nuclear reactions.
The knowledge of these latter is thus crucial for extracting information
on the solar interior from neutrino observations.

In this short review we shall discuss a few items: \\
i) what can be learnt on the Sun from measurement of
the Boron flux?\\
ii) what can be learnt about energy generation in the Sun 
from solar neutrinos? \\
iii) which nuclear physics measurements are now 
crucial for extracting astrophysical information from 
solar neutrino experiments?

\section{The Boron flux: nuclear physics and astrophysics}

\begin{figure}[t]
\begin{center}
\vspace{-1cm}
\includegraphics[width=0.6\textwidth]{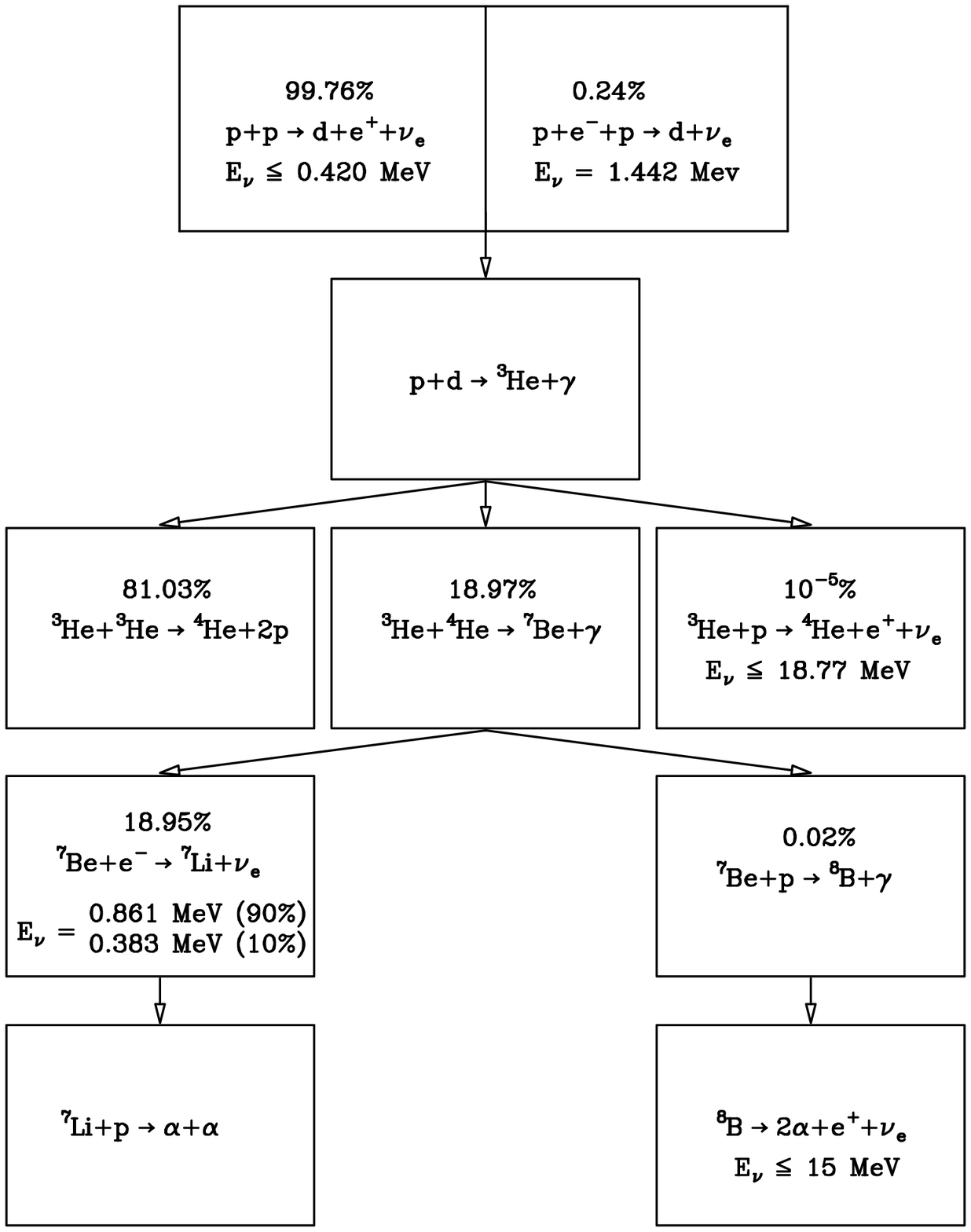}
\end{center}
\vspace{-0.5cm}
\caption[]{
{\bf The pp-chain}. Probabilities of the different branches and the
neutrino energies are indicated.}
\vspace{-5.5cm}
~~~~~~~~~~~~~~~~~~~~~~~~pp-I
\vspace{5cm}

\vspace{-1.3cm}
~~~~~~~~~~~~~~~~~~~~~~~~~~~~~~~pp-II
~~~~~~~~~~~~~~~~~~~~~~~~~~~~~~~pp-III
\vspace{1.cm}
\label{figppchain}
\end{figure}

Among the various branches of the pp-chain, see Fig. \ref{figppchain}, 
 the status of $^8B$ neutrinos, 
the component originating from the pp-III branch, is unique in two
respects: \\
i) experiments as SNO and SuperKAMIOKANDE are
sensitive to these neutrinos only (whereas the signal in 
Chlorine and Gallium radiochemical
experiments is a weighted sum of several
components); \\
ii) the total active neutrino flux from $^8B$ decay
$\Phi_B=\Phi(\nu_e+\nu_\mu+\nu_\tau)$ 
is now a measured quantity.

By combining the final SuperKAMIOKANDE electron
scattering data and the latest SNO charged and neutral current fluxes
one obtains \cite{gf.fogli03}:
\begin{equation}
\Phi_B=5.5 \, (1\pm 7\%)  10^6 cm^{-2} s^{-1} \quad (1\sigma) \,,
\end{equation}
in good agreement with the predictions of
recent SSM calculations, all prior to
this important experimental result, see Table \ref{tabssm}.

\begin{table}[th]
\caption{Predictions of some recent SSM 
calculations compared with experimental result.}
\begin{center}
\renewcommand{\arraystretch}{1.4}
\setlength\tabcolsep{5pt}
\begin{tabular}{lcccc}
\hline\noalign{\smallskip}
                        &  EXP. & \multicolumn{3}{c}{THEORY} \\
                        &  \cite{gf.fogli03}    & BP01 \cite{gf.bp2000}& FRANEC \cite{gf.ciacio} & GARSOM \cite{gf.garsom} \\
\noalign{\smallskip}
\hline
\noalign{\smallskip}
$\Phi_B \, [10^6 cm^{-2} s^{-1}]$              &  $ 5.5 (1\pm 7\%) $ & 5.05 & 5.20 & 5.30 \\
$T \,[10^7 K]$                                 &                     &   1.5696    & 1.569     &  1.57    \\
\hline
\end{tabular}
\end{center}
\vspace{-1cm}
\label{tabssm}
\end{table}

The seven per cent accuracy, which is already remarkable, 
could be possibly improved in the next few years, as a consequence 
of higher statistics and better experimental techniques.


\begin{table}[t]
\vspace{-0.3cm}
\caption{{\bf Neutrino fluxes, central solar temperature and solar model
inputs.} Each column contains
the logarithmic partial derivates of the neutrino fluxes and $T$ with
respect to the parameter  shown at the top of the column. All the values
have been computed with FRANEC including
elemental diffusion. 
Generally there is good agreement with \cite{gf.jnb}:
in parenthesis we report the corresponding values from \cite{gf.jnb} when
the difference exceeds 10\%.}
\begin{center}
\renewcommand{\arraystretch}{1.4}
\setlength\tabcolsep{3pt}
\begin{tabular}{l|ccccc|ccccc|}
\hline\noalign{\smallskip}
       &\multicolumn{5}{c|}{Nuclear} & \multicolumn{5}{c|}{Astrophysical} \\
$   $ & $S_{11}$  &  $S_{33}$  &  $S_{34}$    &  $S_{17}$  & $S_{1\, 14}$ &
      $lum$  & $comp$  &  $opa$  &  $age$ &  $dif$ \\
\hline
pp & 0.114   & 0.029 & -0.062 & 0 &- 0.019 & 
      0.73 & -0.076 &-0.12 & -0.088  & -0.02 \\
   &     {\em (0.14)} &&&&&&&& {\em (-0.07)} & \\

Be & -1.03 & -0.45 & 0.87 & 0 & -0.027   &
      3.5 & 0.60 & 1.18 & 0.78  &0.17 \\
   & &&&& {\em (-0.00)} &&&& {\em (0.69)} & \\

B & -2.73 & -0.43 & 0.84 &1 &-0.02  & 
      7.2 & 1.36 & 2.64& 1.41 & 0.34 \\
  &&&&& {\em (+0.01)} &&&&&\\

N & -2.59 & 0.019 & -0.047 & 0 & 0.83 &
     5.3 & 1.94 & 1.82 &  1.15& 0.25 \\
  & &&&&&&&& {\em (1.01)} &\\

O & -3.06 & 0.013&  -0.038  & 0 & 0.99 &
     6.3& 2.12 & 2.17 & 1.41 & 0.34 \\
  && {\em (0.02)} & {\em (-0.05)} &&&&&&&\\

\hline
$T$ & -0.14 & -0.0024 & 0.0045 & 0 & 0.0033 &
      0.34 & 0.078 & 0.14 & 0.083 & 0.016 \\
\hline
\end{tabular} 
\end{center}
\vspace{-0.5cm}
\label{tableggi}
\end{table}


The Boron flux, as well as the other components,
 depends on several nuclear physics and astrophysical 
inputs $X$, see e.g. \cite{gf.report,gf.what}. 
Scaling laws give the variation of fluxes with respect to SSM calculations 
when the input parameter $X$ is slightly changed from  the SSM value
$X_{SSM}$:
\begin{equation}
\Phi_i=\Phi_i^{SSM} \, (X/X^{SSM})^{\alpha_X}
\end{equation}

The power law coefficients $\alpha_X$, 
derived with FRANEC models including
diffusion, are collected in Table \ref{tableggi}. 
Generally there is excellent agreeement between our
calculated values and those in ref. \cite{gf.jnb}.

For the Boron flux one has:
\begin{eqnarray}
\label{eqleggi}
\Phi_B &= & \Phi_B^{SSM} \, s_{33}^{-0.43} \, s_{34}^{0.84} \,s_{17}^{-1}
\, s_{e7}^{-1} \, s_{11}^{-2.7} \nonumber \\
       & \cdot  & lum^{7.2} \, comp^{1.4} \,opa^{2.6} \,age^{1.4} \,dif^{0.34} 
\end{eqnarray}
where for each parameter $x=X/X^{SSM}$. The first line 
contains the nuclear physics parameters 
($S_{ij}$ are the
astrophysical
factors at zero energy for nuclear reactions $i+j$),
and the second line groups 
the astrophysical inputs: \\
-$lum = (L/L_\odot)$ expresses the sensitivity to the solar luminosity;\\
-$comp = (Z/X) /(Z/X)^{SSM}$ accounts for the metal content of the solar
photosphere; \\
-$age = (t/t_\odot)$ expresses the sensitivity to the solar age;\\
-$opa$ and $dif$ are uniform scaling parameters with respect to the
opacity tables and the diffusion coefficients used in SSM calculations.

Eq. \ref{eqleggi} shows that from a flux measurement 
 one can learn astrophysics if nuclear physics is known well enough.

\subsection{The uncertainty budget on $\Phi_B$}

In Table \ref{tabBbudget} we present the uncertainty budget, including
errors on the inputs and the propagated effect on $\Phi_B$. 
This table deserves several comments.\\

i)In the last few years there has been a significant progress
in the experimental study of low energy nuclear reactions. 
In particular  LUNA performed in the underground Gran Sasso laboratory
has measured the $^3He+^3He$ $\rightarrow^4He+2p$ down to
solar energies, avoiding extrapolations.
This resulted in a reliable determination of $S_{33}$ with 6\% accuracy 
\cite{gf.luna33}.\\

ii) Concerning the reaction
$^7Be+p\rightarrow^8B+\gamma$,
until a few years ago the uncertainty on the astrophysical factor was at
the level of 10-15\%:
two reviews, published in 1998, recommended  
$S_{17}=19^{+4}_{-2}$ eVb \cite{gf.adelberger}
and $S_{17}=21\pm2$ eVb \cite{gf.nacre}.
Several new experiments were performed in the last few years, see
Table \ref{tabs17}.

\begin{table}[hb]
\caption{{\bf Uncertainties budget on Boron neutrino flux}. 
All the values are at $1\sigma$ level.}
\begin{center}
\renewcommand{\arraystretch}{1.4}
\setlength\tabcolsep{5pt}
\begin{tabular}{lcc}
\hline\noalign{\smallskip}
Source & ~~~$\Delta X/ X \, (\%)$ & $\Delta \Phi_B/\Phi_B \, (\%)$ \\
\noalign{\smallskip}
\hline
\noalign{\smallskip}
$S_{33}$ & 6$^*$ & 3 \\
$S_{34}$ & 9$^\dag$ & 8 \\
$S_{17}$ & 5~ & 5 \\
$S_{e7}$ & 2$^\dag$  & 2 \\
$S_{11}$ & 2$^\dag$  & 5 \\
\hline
$lum$  & 0.4$^\ddag$  & 3 \\
$comp$ & 7~  &  10 \\
$opa$     & 2.5$^{**}$  &  7 \\
$age$     & 0.4$^\ddag$  & 0.6  \\
$dif$   & 10~  & 3 \\

\hline
Exp.    &        & 7  \\
\hline
\multicolumn{3}{l}{$^*$ from \cite{gf.luna33}; $\, ^\dag$ from \cite{gf.adelberger};
$\, ^\ddag$ from \cite{gf.bp95}; $\, ^{**}$ from \cite{gf.bp92}  }\\
\end{tabular}
\end{center}
\label{tabBbudget}
\end{table}

Quite recently new measurements
have been presented \cite{gf.junghans}, ranging from
$E_{cm} =116$ to 2460 keV, and
incorporating several improvements over the previously published
experiment \cite{gf.junghanspr}. 
This new measurement yields $ S_{17}=22.1\pm 0.6(expt)\pm 0.6(theor)$ eVb
based on data from
$E_{cm} =116$ to 362 keV.
The central value is obtained from the theoretical shape 
predicted by 
Descouvemont and Baye \cite{gf.db}.
The theoretical error is based on the fit of 12 different
theories to the low energy data.

In addition Junghans et al. \cite{gf.junghans}
compare the results
of all ``modern'' direct experiments, by using the same theoretical curve
in fitting the data. They find: 
\begin{eqnarray}
 S_{17} & = &21.4 \pm 0.5 \, eVb \quad   \, E_{cm}< 425\, KeV  \quad  \chi^2/d.o.f.=1.2\\
 S_{17} & = &21.1 \pm 0.4 \, eVb \quad   \, E_{cm}< 1200\, KeV  \quad  \chi^2/d.o.f.=2.1 \, .
\end{eqnarray}	  
The fit at the low-energy region is quite good, whereas
the wide-range suggests that some of the
uncertainties may be underestimated. 
In conclusion, they recommend as ``best'' value:
\begin{equation}
S_{17}=21.4\pm 0.5(expt)+0.6(theor)\, eV b \quad   (1\sigma).
\end{equation}

We remind however that the low-energy global fit is dominated by the data
of ref. \cite{gf.junghans}, all other ``modern'' experiments
yielding somehow lower $S_{17}$  values. 
Indirect methods for determining
$S_{17}$ (Coulomb dissociation, heavy ion transfer and breakup) also
suggest a somehow smaller value.
In conclusion it looks that a 5\% accuracy on $S_{17}$ has been
reached.

\begin{table}[th]
\caption{{\bf Results on $S_{17}$ from direct capture experiments.}}
\begin{center}
\renewcommand{\arraystretch}{1.4}
\setlength\tabcolsep{5pt}
\begin{tabular}{ll}
\hline\noalign{\smallskip}
 $S_{17}(0) \, [eV b]$ & Ref. \\
\noalign{\smallskip}
\hline
\noalign{\smallskip}
$19^{+4}_{-2}$ & Adelberger et al. compilation \cite{gf.adelberger}\\
$21 \pm 2 $  & NACRE compilation \cite{gf.nacre}\\
\hline
$20.3 \pm 1.2$  & Hass et al. \cite{gf.hass}\\
$18.8  \pm 1.7 ^\dag$  & Hammache et al. \cite{gf.hammache}\\
$ 18.4 \pm 1.6$  & Strieder et al. \cite{gf.strieder}\\
$21.2 \pm 0.7 $  & Baby et al. \cite{gf.baby}\\
$22.1 \pm 0.6 $ &   Junghans et al. \cite{gf.junghans}\\
\hline
\multicolumn{2}{l}{$^\dag$ theoretically uncertainty included}
\end{tabular}
\vspace{-1cm}
\end{center}
\label{tabs17}
\end{table}

~\\
iii) Concerning the metal fraction $Z/X$, by
using the values reported in \cite{gf.grevesse} and by propagating
the individual uncertainties one finds \cite{gf.metal}:
\begin{equation}
Z/X_\odot=0.0233 \pm 0.0166  \quad (1\sigma).
\end{equation}
This 7\% uncertainty is similar to that estimated in
\cite{gf.bp95}, on the grounds of the spread among the $Z/X$ estimates
published from 1984 until 1993. \\

iv) With regards to the diffusion coefficient, we assume a 10\%
uncertainty
on the grounds that larger variations would spoil the agreement with
helioseismic results \cite{gf.diff}.

Our uncertainty budget is  similar to that presented in
\cite{gf.jnbtable}, the main difference regarding 
the error on $S_{17}$ (Bahcall refers to the
Adelberger estimate).
Also, concerning the effect of diffusion, a more conservative 
estimate is adopted in \cite{gf.jnbtable}, where the uncertainty
is obtained by comparing models with and without diffusion.

In conclusion the accuracy on the measured Boron neutrino flux
is already comparable to astrophysical uncertainties of the solar
model. The 9\% error of $S_{34}$ is presently the
main source of uncertainty for extracting information on
solar properties from the measurement of the $^8B$ neutrino flux.
In this respect, the planned new measurement of $^3He+^4He$
cross section by LUNA at Gran Sasso is most important.

\subsection{The central solar temperature}

As well known Boron neutrinos are  an excellent solar thermometer, since
the produced flux depends on a high ($\simeq 20$) power  
of the temperature near the solar center $T$. 
It is now time to rediscuss this possibility of exploring the
solar interior since
the produced flux has been measured.

\begin{figure}[ht]
\begin{center}
\includegraphics[width=0.4\textwidth,angle=90]{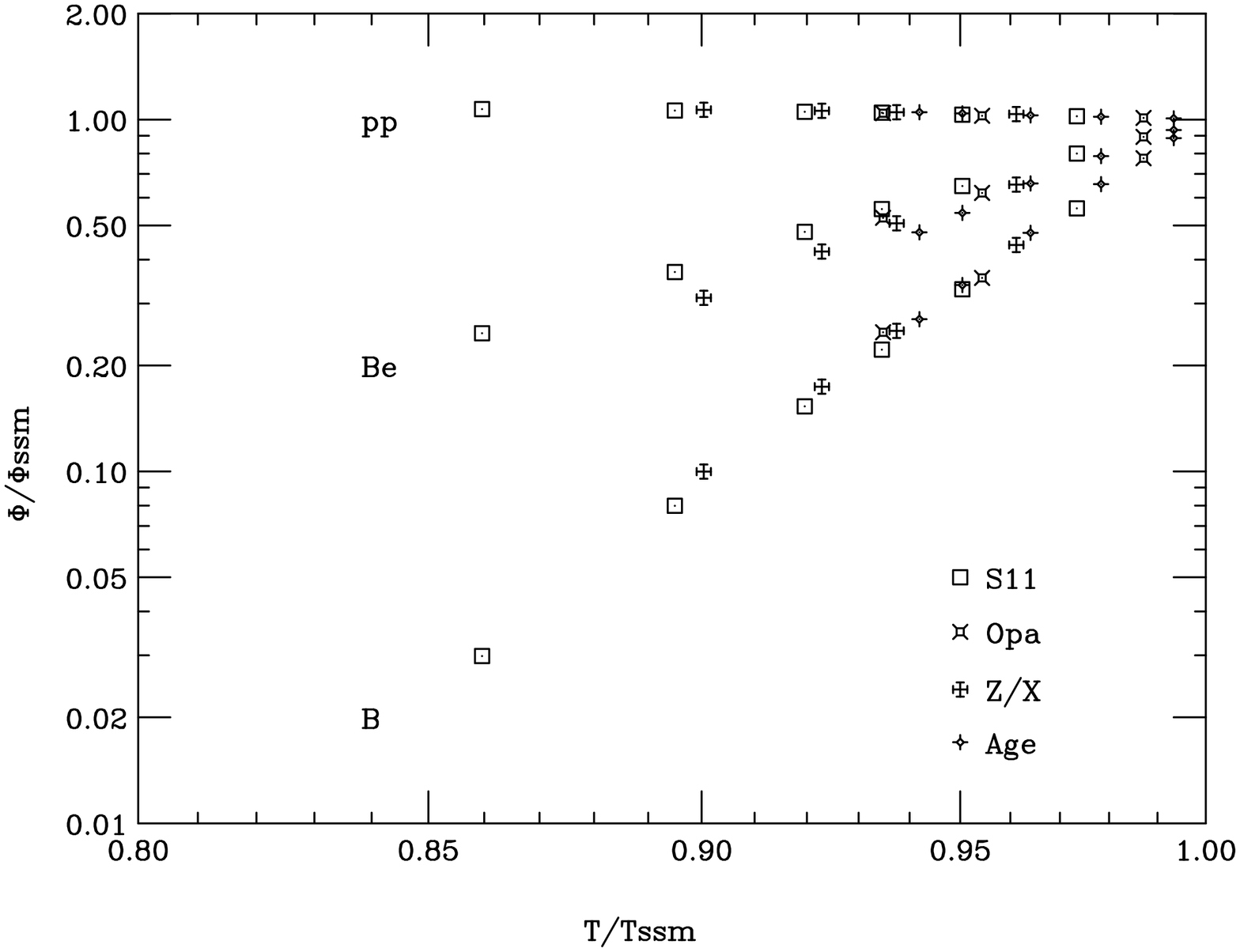}
\includegraphics[width=0.6\textwidth]{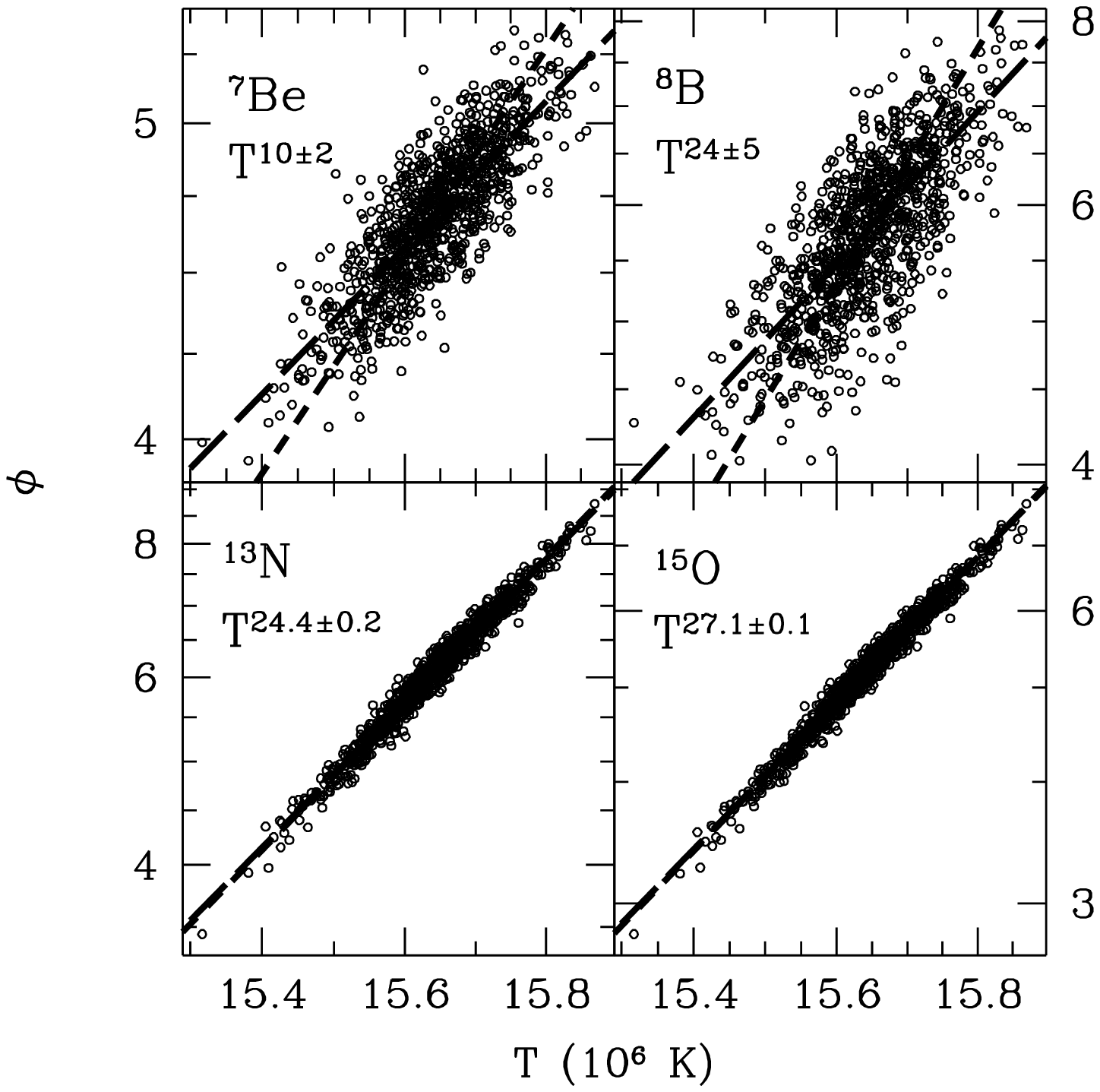}
\end{center}
\vspace{-1.5cm}
\caption[]{The behaviour of pp, Beryllium, and Boron neutrinos as a
function of 
the central temperature $T$ when varying different input parameters, 
a) from \cite{gf.future},
b) from \cite{gf.bu}.}
\vspace{-5.5cm}
~~~~~~~~~~~~~~~~~~~~~~~~b)
\vspace{5cm}

\vspace{-12cm}
~~~~~~~~~~~~~~~~~~~~~~~~a)
\vspace{11.5cm}
\label{figtemp}
\end{figure}

We remind that $T$ is not an independent quantity, its value
being the result of the physical and chemical 
properties of the star.
Actually, the various inputs to $\Phi_B$ in eq. \ref{eqleggi}
can be grouped according to their effect on $T$. 
All nuclear inputs but $S_{11}$ only determine the weight of the
different branches ppI/ppII/ppIII without changing solar structure
and temperature. On the other hand, to a large extent the effect of the
others can be reabsorbed into a variation of the central
solar temperature, 
almost independently on the way we use
to vary it, see 
Fig. \ref{figtemp}.

\begin{table}[th]
\caption{{\bf Central solar temperature and B neutrino flux.}}
\begin{center}
\renewcommand{\arraystretch}{1.4}
\setlength\tabcolsep{5pt}
\begin{tabular}{llll}
\hline\noalign{\smallskip}
Source & $\partial ln T/ \partial lnX$  &  $\partial ln \Phi_B/ \partial lnX $ &  \\
       &  $\beta$         &        $\alpha$        &   $\alpha/\beta$       \\
\noalign{\smallskip}
\hline
\noalign{\smallskip}
$S_{33}$ & 0 & -0.43 & \\
$S_{34}$ & 0 & 0.84 & \\
$S_{17}$ & 0 & 1 &  \\
$S_{e7}$ & 0 & -1 & \\
\hline
$S_{11}$ & -0.14 & -2.7 & 19 \\
$lum$  & 0.34 & 7.2 & 21 \\
$comp$ & 0.08 & 1.4  &17\\
$opa$     & 0.14 & 2.6 & 19 \\
$age$   & 0.08 & 1.4 & 17 \\
$dif$   & 0.016 & 0.34 & 21 \\

\hline
\end{tabular}
\end{center}
\vspace{-0.3cm}
\label{tabtemp}
\end{table}

This is shown more quantitatively 
in Table \ref{tabtemp}, where one sees 
a near constancy of $\alpha/ \beta = \frac{\partial ln\Phi}{\partial lnX} \,
\frac{\partial ln X}{\partial ln T}$ .
the values in the last column. 
In  summary we can write:
\begin{equation}
\Phi_B=\Phi_B^{SSM} (T/T^{SSM})^{20} s_{nuc}
\end{equation}
where the coefficient $20$ is an average of the calculated
$\alpha/\beta$ (see Table \ref{tabtemp}) 
and $s_{nuc}=s_{33}^{-0.43} \, s_{34}^{0.84} $ $\, s_{17}^{-1} \, s_{e7}^{-1} $

The agreement of theory and experiment on the Boron neutrino flux
means thus that we can take $T=1.57\, 10^7 \, K$ 
as the solar temperature in the region where Boron 
neutrinos are produced  \cite{gf.what}. 
The present  experimental uncertainty on $\Phi_B$ (7\%)
and the error on $s_{nuc}$ (10\%)
yield:
\begin{equation}
\Delta T/T= 0.6\%
\end{equation}
where the main uncertainty arises from $S_{34}$. In other words,
a crucial prediction of SSM has been verified with neutrinos
with an accuracy better than 1\%.

\begin{figure}[t]
\begin{center}
\vspace{-2cm}
\includegraphics[width=0.7\textwidth]{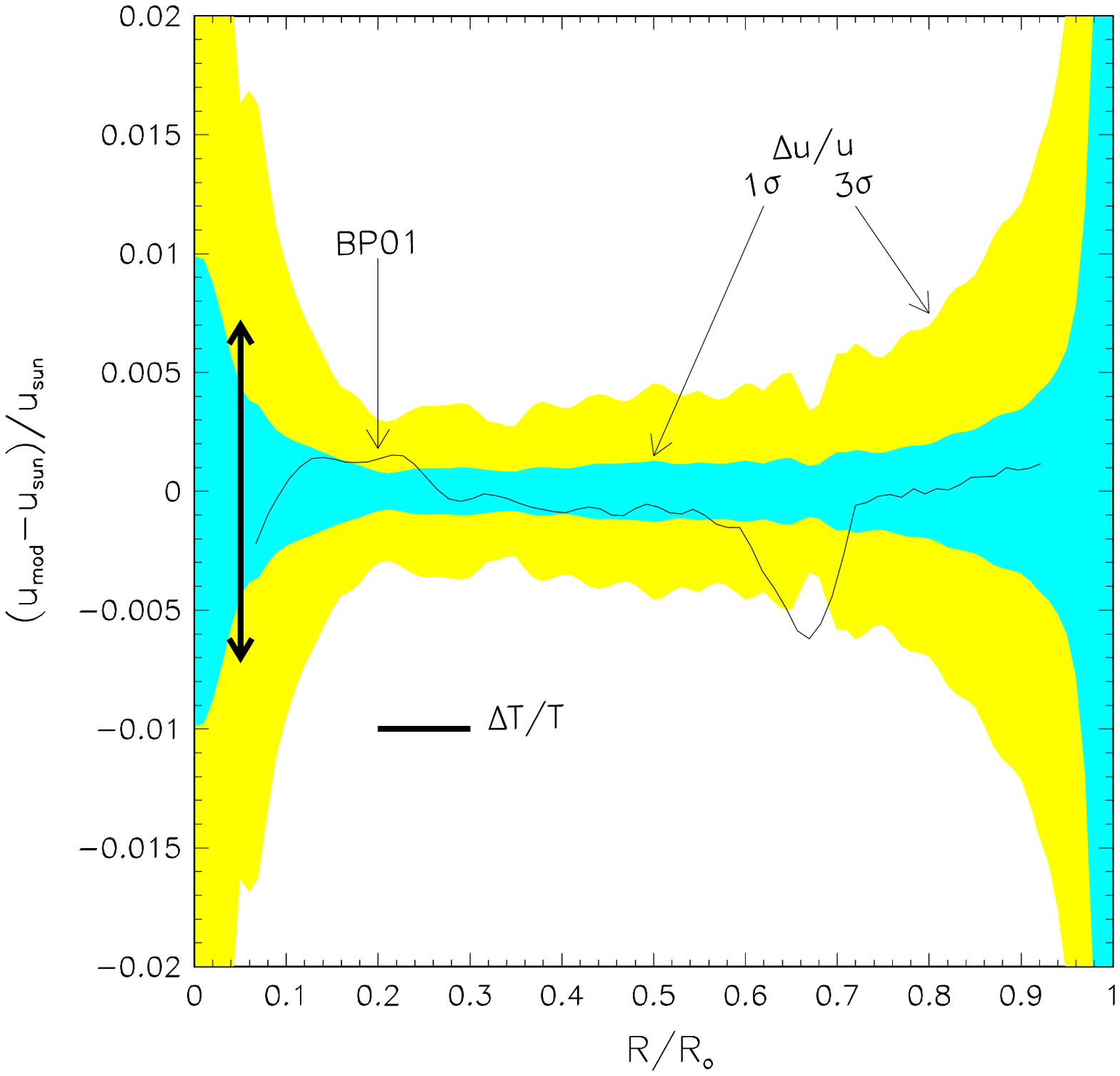}
\end{center}
\vspace{-2.5cm}
\caption[]{The dark (light) shaded area corresponds to the 
$1\sigma \, (3\sigma)$ uncertainty on helioseismic determination of
squared isothermal sound speed  $U=P/\rho$ \cite{gf.elio}. 
The relative difference between  the SSM
prediction \cite{gf.bp2000}
and the helioseismic data is also shown (thin line). 
The present 
uncertainty on the the solar temperature as derived by
the measurement of Boron neutrino flux is indicated with the  thick line.}
\label{figelio}
\end{figure}

A comparison with helioseismology is useful. 
Helioseismology allows us to look into the deep interior 
of the Sun, see e.g. \cite{gf.elios,gf.bp2000}. 
The highly  precise  measurements of 
frequencies and  the tremendous number of measured lines enable us to extract both
the properties of the convective envelope (depth, helium abundance) and 
the sound speed along the solar profile   with high accuracy. 
This latter quantity  is determined to the level of about 0.15\%
in a large portion of the Sun. The accuracy degrades to about 1\%
near the center, see Figure \ref{figelio}.

From helioseismic observations one  cannot determine 
directly the temperature of the solar interior, as  one cannot determine
the temperature of a gas from the knowledge of the sound speed unless the 
chemical composition is known. 

However, it is possible to obtain the range of helioseismic
allowed values of the central temperature $T$, by
selecting those solar models which are consistent with 
seismic data. 
More specifically, in ref. \cite{gf.eliostc}
the central temperature $T_{helios}$ has been determined  as that of the
model which gives the best fit to the seismic data.
The uncertainty, 
$\Delta T_{helios}$, corresponds to the range spanned by models
consistent with these data. 
This results in
$T_{helios}=1.58 \, 10^7 K$, in good agreement with the SSM
predictions, and  
 $\Delta T/T_{helios}=0.5 \%$ at $1\sigma$.

The neutrino result, which is much more direct, is an excellent
confirmation of helioseismic inferences. The accuracies of the
two methods are already comparable and one can expect neutrinos
to become more accurate as better flux and cross sections measurements
will be available.

\subsection{The sun as laboratory}

In the next few years one can envisage a measurement of the solar
temperature near the center with an accuracy of order
of 0.1 per cent, as the result of progresses in 
neutrino  and nuclear physics. 
This can be relevant in several respects: \\
1) it will provide a new challenge to SSM calculations;\\
2) it will allow a determination of the metal content in the solar
interior, which has important consequences on the history 
of the solar system \cite{gf.metal};\\
3) one can find constraints (or surprises, or even discoveries)
on several issues, as e.g. axion emission from the sun, 
the physics of extra dimensions, dark matter...

All this shows that the Sun is really becoming a laboratory for
astrophysics and fundamental physics.

\section{pp-chain, CNO cycle or what else?}

According to our understanding of the Sun, most of its power
originates from the pp-chain, with a minor contribution
($\approx 1\%$) from the CNO cycle. 
Although this is theoretically well grounded, an experimental
verification is clearly welcome. 

From the theoretical point of 
view, solar model predictions for CNO neutrino
fluxes are not precise because
the CNO fusion reactions are not as well studied as the pp reactions, 
see Table \ref{tabCNObudget}.
Also, the Coulomb barrier is higher for the CNO reactions, implying a
greater sensitivity to details of the solar model.

The principal error source is $S_{1\,14}$, the astrophysical
S-factor of the slowest reaction in the CNO cycle,
$^{14}N(p,\gamma)^{15}O$.
At solar energies this reaction is dominated by a
sub-threshold resonance at -504 keV, whereas at energies higher than 100
keV it is dominated by the 278 keV resonance, with transitions to the
ground-state of $^{15}O$ or to the excited states at 
energies of 5.18 MeV, 6.18 MeV and 6.79 MeV.

According to Schroeder et al. \cite{gf.schreder}, who measured
 down to 200 keV, the
main contribution to the total S-factor at zero energy comes from
the transitions to the ground state of $^{15}O$ 
and to its excited state at $E_x = 6.79$ MeV. In particular,
they give $S_{1\,14} = 3.20 \pm 0.54$ keVb. 
Angulo et al. \cite{gf.angulo}
reanalyzed Schroeder experimental data using an R-matrix model.
They obtained
$S_{1\,14} = 1.77 \pm 0.20$ keVb, a factor 1.7 below that of 
\cite{gf.schreder}.
LUNA at Gran Sasso will soon clarify this uncertainty: high accuracy data
have been taken down to 180 keV and the preliminary 
results \cite{gf.broggini} show the possibility of LUNA to discriminate
between the two different extrapolation.

\begin{table}[th]
\caption{{\bf Uncertainties budget on CNO neutrino flux}. All
the values are at $1\sigma$ level. See Table 3 for estimates of 
$\Delta X/X$.}
\begin{center}
\renewcommand{\arraystretch}{1.4}
\setlength\tabcolsep{5pt}
\begin{tabular}{llll}
\hline\noalign{\smallskip}
Source & $\Delta X/ X \, (\%)$ & $\Delta \Phi_N/\Phi_N$ (\%)& $\Delta
\Phi_O/\Phi_O$ (\%)\\
\noalign{\smallskip}
\hline
\noalign{\smallskip}
$S_{33}$ & 6 & 0.1 & 0.08 \\
$S_{34}$ & 9 & 0.4 & 0.3\\
$S_{17}$ & 5 & 0 & 0\\
$S_{e7}$ & 2  & 0 & 0\\
$S_{11}$ & 2  & 5 & 6 \\
$S_{1\,14}$ & $^{+11 \, \, \dag}_{-46} $   & $^{+9}_{-38}$   &  $^{+11}_{-46} $\\
\hline
$lum$  & 0.4  & 2 & 3 \\
$comp$ & 7  &  13 & 15 \\
$opa$     & 2.5  &  5 & 5 \\
$age$   & 0.4 & 0.46 & 0.56 \\
$dif$   & 10  & 3 & 3\\
\hline
\multicolumn{4}{l}{$^\dag$ from \cite{gf.adelberger}}
\end{tabular}
\vspace{-0.5cm}
\end{center}
\label{tabCNObudget}
\end{table}

From a global analysis of solar neutrino experiments \cite{gf.jnbcno}, 
Bahcall et al. derived  an upper limit $(3\sigma)$
of 7.8\% (7.3\% including the  KamLAND measurements) to the fraction of
energy that the Sun produces via the CNO fusion cycle.

As mentioned at the beginning of this 
subsection, the important underlying questions
are: is the Sun fully powered by nuclear reactions?
Are there additional energy losses, beyond photons  and neutrinos?

The idea that the Sun shines because of nuclear fusion
reactions can be tested accurately by comparing the observed photon
luminosity of
the Sun $L_\odot (\gamma)$ 
with the luminosity inferred from measurements of solar
neutrino fluxes, $L_\odot (\nu )$.
In fact for each fusion of four proton into a Helium nucleus
\begin{equation}
\label{pp}
4p+2e^- \rightarrow ^4He+2\nu_e
\end{equation}
an energy $Q=26.73$ MeV is released together with two
neutrinos. If one determines from experiments the 
total neutrino production rate one is also
determining the energy production rate in the Sun by
means of (\ref{pp}), see sect. 2.1 of \cite{gf.report}.

Bahcall and Pena-Garay \cite{gf.roadmap}
performed a global analysis of all the available solar and reactor
data to determine the $1\sigma \, (3\sigma)$ allowed range for
$L_\odot (\nu )$.
For the ratio to the accurately measured
photon luminosity, they find:
\begin{equation}
\frac{L_\odot (\nu)} {L_\odot(\gamma)}= 
1.4 ^{+0.2} _{-0.3} \left (^{+0.7}_{-0.6} \right ) 
\end{equation}

At $1\sigma$ the luminosity of the Sun as inferred from
neutrinos is thus determined to about 20\%. 
Note that at $3\sigma$ the neutrino-inferred solar luminosity 
can be as large as
(as small as) 2.1 (0.8) the precisely measured photon-luminosity.

A $^7Be$ solar neutrino experiment accurate to 5\% 
could improve this determination to about 13\%. 
The global combination of a $^7Be$ experiment, 
plus a p-p experiment, plus the
existing solar data and three years of KamLAND would make possible a 
really precise determination of the solar energy produced by 
nuclear reactions, see \cite{gf.roadmap}.

\section*{Acknowledgment}
We are grateful to E. Lisi and A. Marrone for providing
us with the plots in Fig.~\ref{figsno}, and to C. Broggini for
useful comments.

%

\end{document}